\documentclass[aps,prd,a4paper,nofootinbib,
preprintnumbers,superscriptaddress,showpacs,showkeys,twocolumn]{revtex4-2}

\usepackage{amsmath}
\usepackage{amssymb}
\usepackage{bm}
\usepackage{graphicx}
\usepackage{color}

\newcommand{\be}{\begin{equation}}
\newcommand{\ee}{\end{equation}}
\newcommand{\ba}{\begin{eqnarray}}
\newcommand{\ea}{\end{eqnarray}}

\begin{document}

\title{Memory effects on energy loss and diffusion
	of
	heavy quarks in the quark-gluon plasma}

\author{Marco Ruggieri}\email{ruggieri@lzu.edu.cn}
\affiliation{School of Nuclear Science and Technology, Lanzhou University, 222 South Tianshui Road, Lanzhou 730000, China}

\author{Pooja}
\affiliation{School of Physical Sciences, Indian Institute of Technology Goa, Ponda-403401, Goa, India}

\author{Jai Prakash}
\affiliation{School of Physical Sciences, Indian Institute of Technology Goa, Ponda-403401, Goa, India}

\author{Santosh K. Das}
\affiliation{School of Physical Sciences, Indian Institute of Technology Goa, Ponda-403401, Goa, India}


\begin{abstract}
We study the dynamics of heavy quarks in a 
thermalized quark-gluon
plasma with a time-correlated thermal noise, $\eta$. 
In this case it is said that
$\eta$ has memory.
We use an integro-differential
Langevin equation in which the memory enters  via the
thermal noise and the dissipative force.
We assume that the time correlations of the noise decay
exponentially on a time scale, $\tau$, that we treat as a
free parameter. 
We compute the effects of $\tau\neq 0$ on 
the thermalization time of the heavy quarks,
on their momentum broadening 
and on the nuclear modification factor. We find that overall
memory slows down the momentum evolution of heavy quarks:
in fact, transverse momentum broadening
and the formation of $R_{AA}$ are slowed down by memory 
and the thermalization time of the heavy quarks become 
larger. The potential impact on other observables is
discussed briefly.  
\end{abstract}

\pacs{12.38.Aw,12.38.Mh}

\keywords{Relativistic heavy ion collisions, 
	heavy quarks, Langevin equation, processes with memory, 
nuclear modification factor, quark-gluon plasma}

\maketitle

\section{Introduction}

A hot and  dense phase of nuclear matter, the quark-gluon plasma (QGP), is
expected to form  in the ultra-relativistic heavy-ion collisions at Relativistic Heavy-ion Collider (RHIC) and the Large Hadron Collider (LHC) energies. 
Probing and characterizing  the bulk
properties of QGP is a field of high contemporary interest. 
Heavy quarks (HQs)~\cite{Prino:2016cni,Andronic:2015wma,Rapp:2018qla,Aarts:2016hap,Cao:2018ews,Dong:2019unq,Xu:2018gux,GolamMustafa:1997id,Uphoff:2011ad,Song:2015sfa,Cao:2016gvr,Plumari:2017ntm} such as charm and beauty
are considered as good probes of the system produced in high-energy nuclear collisions. 
In fact, 
they are produced in the very
early stage due to the hard partonic scatterings
on a time scale $\tau=O(1/m)$ where $m$ is
the rest mass of the quark.
Due to their large mass and low abundance, they
can propagate in the QGP bringing
almost no disturbance to it.
Consequently, they act as good probes that can
experience the whole evolution of the system
created in the collisions, from the very early
stage up to the hadronization stage.

The standard approach to study the HQ dynamics in the QGP is following their position and momentum 
evolution by means of the Langevin 
equations~\cite{rappv2,rappprl,Gossiaux:2008jv,Das:2010tj,Alberico:2011zy,Lang:2012cx,He:2012df,He:2013zua,Das:2013kea,Cao:2015hia,Das:2016cwd,Xu:2017obm,Katz:2019fkc,Li:2020umn} (see also \cite{Schmidt:2014zpa})
as well as relativistic kinetic theory 
\cite{Uphoff:2011ad,Song:2015sfa,Cao:2016gvr,Plumari:2017ntm,Das:2013kea,Scardina:2017ipo,Uphoff:2012gb,Song:2015ykw,Nahrgang:2014vza}.
In the approaches based on the Langevin equation,
the thermal noise, $\eta$, is usually
treated as standard Wiener process,
thus without correlations
in time. In this work we relax this approximation and analyze 
the case in which $\eta$ is time-correlated;
this class of stochastic processes is called
a process with memory.

The prototype of
Langevin equation 
that we consider in this work is
\begin{equation}
	\frac{dp}{dt} = -\int_0^t \gamma(t-s)p(s) ds + \eta(t),
	\label{eq:L1i}
\end{equation}
where $p$ is the momentum of the particle, 
$\eta$ is the stochastic term that 
models the thermal noise, while 
the integral term on the right hand side is the dissipative force. In previous studies 
the latter is 
replaced by $-\gamma p$ where $\gamma$ is the drag coefficient:
this replacement follows 
from the Fluctuation-Dissipation
Theorem (FDT) when $\eta$ has no time correlations.

Our goal is to analyze the motion of heavy quarks in a
quark-gluon plasma, when the correlations of the thermal
noise do not decay instantaneously: instead, we assume that
these correlations decay over a specific time scale that we
call the memory time, $\tau$. Hydrodynamic 
fluctuations ~\cite{Kapusta:2011gt,Murase:2016rhl},
diffusion in the evolving Glasma
\cite{Liu:2019lac,Liu:2020cpj,Khowal:2021zoo,Mrowczynski:2017kso,Ipp:2020nfu,Carrington:2020sww,Ruggieri:2018rzi,Boguslavski:2020tqz,Ipp:2020mjc,Sun:2019fud}, diffusion of electric charge~\cite{Kapusta:2017hfi}, dilepton yields~\cite{Schenke:2006uh} and the electric conductivity
of the quark-gluon plasma ~\cite{Hammelmann:2018ath} 
are some of the physical problems where memory can play a role;
for these $\tau$ lies in between
$0.1$ and $10$ fm/c. In this study 
we treat $\tau$ as a free parameter
and study its effect on a few physical quantities,
namely the momentum broadening of heavy quarks and 
on the nuclear modification factor, $R_{AA}$.
For the sake of simplicity we 
consider the interaction 
of heavy quarks with a thermalized
quark-gluon plasma at a fixed, constant temperature $T$;
the diffusion coefficients that we use in the calculations
are that obtained by perturbative QCD (pQCD) for high $T$, 
and by a quasiparticle model (QPM) for low $T$,
while the dissipative kernel is related to the
thermal noise by the Fluctuation-Dissipation Theorem
(FDT).

We anticipate the main result, namely that 
memory delays the dynamics of the heavy quarks in the
QGP: we show this by studying the thermalization time,
the momentum broadening and the time evolution of $R_{AA}$. 
The latter in particular 
can be potentially of interest for the phenomenology
of heavy quarks in the QGP, due to the fact that 
the slower evolution of  $R_{AA}$ would require the use
of larger diffusion coefficients in order to
reproduce the experimental data and this in turn would
require stronger interactions of the heavy quarks with the
bulk, potentially leading to a larger $v_2$.

The plan of the article is the following: in section
\ref{sec:iianalyt} we present the calculations
for the equilibration time and the momentum broadening
for the Brownian motion with memory in the nonrelativistic
limit; 
in section \ref{sec:laimple} we discuss the numerical
implementation of the Langevin equation with 
an integral kernel, while in section IV we present our results.
Finally, in section V we draw our conclusions.

\section{Nonrelativistic limit\label{sec:iianalyt}}
For the sake of illustration, we consider here a simple
one-dimensional motion of a heavy particle with mass $m$
in the nonrelativistic limit.
Most of the calculations presented here have been
obtained firstly in \cite{Ruggieri:2019zos},
where a gaussian correlator has been considered. Here we
consider an exponential correlator instead,
therefore we skip many details that have been given in 
\cite{Ruggieri:2019zos}, and limit ourselves to write explicit
results that stand for the exponential correlator.

The Langevin equation for momentum $p$ is
\begin{equation}
	\frac{dp}{dt} = -\int_0^t \gamma(t-s)p(s) ds + \eta(t),
	\label{eq:L1}
\end{equation}
where $\eta$ is the stochastic term that 
models the noise, while 
the integral term on the right hand side is the dissipative force.

The formal solution of Eq.~\eqref{eq:L1} can be obtained
by means of Laplace transforms, namely
\begin{equation}
	p(t) =\frac{1}{2\pi i}\int_{\sigma-i\infty}^{\sigma+i\infty}\!\frac{p_0 + \Xi(s)}{s+\Gamma(s)}e^{s t}ds,\label{eq:solution1}
\end{equation}
where $p_0=p$ at $t=0$, and $\Gamma$ and $\Xi$ denote the Laplace transforms of the dissipative kernel and of the noise respectively;
the integral is understood on a Bromwich contour that leaves all the singularities of the integrand on its left side.

We assume that $\eta$ is a gaussian random variable
with correlators given by
\begin{equation}
	\langle\eta(t_1)\eta(t_2)\rangle =   2{\cal D} f(t_1-t_2).\label{eq:corr3}
\end{equation}
Moreover, we assume that $\eta$ represents the thermal noise
in a thermalized bath at temperature $T$.
From the FDT we have then
\begin{equation}
	\gamma(t,s) = \frac{1}{m T}\langle
	\eta(t) \eta(s)
	\rangle=2\gamma f(t-s),
	\label{eq:corr4bis}
\end{equation}
with
\begin{equation}
\gamma=\frac{\mathcal{D}}{m T}.
\label{eq:varrhDEDO}
\end{equation}

In order to analyze
the thermalization time and the momentum broadening
of the heavy particle 
we need to evaluate the following averages:
\begin{eqnarray}
&&	\left\langle  p(t) \right\rangle =
 p_0{\cal G},\label{eq:17}\\
&&	\sigma_p\equiv \left\langle ( p(t)-\langle p\rangle)^2 \right\rangle  =   {\cal J},\label{eq:18}
\end{eqnarray}
where we have put
\begin{equation}
	{\cal G} = \frac{1}{2\pi i}\int_{\sigma-i\infty}^{\sigma+i\infty}\!\frac{1}{s+\Gamma(s)}e^{s t}ds,\label{eq:solution1aam}
\end{equation}
that describes momentum randomization due to the propagation of the particle in the bath, and
\begin{equation}
	{\cal J} = {\cal L}^{-1}
	\left[
	\frac{\langle\Xi(s_1)\Xi(s_2)\rangle}{(s_1 + \Gamma(s_1))(s_2 + \Gamma(s_2))}
	\right](t,t),\label{eq:solution1anna}
\end{equation}
where  ${\cal L}^{-1}[h](t_1,t_2)$ is the two-dimensional inverse Laplace transform of $h(s_1,s_2)$
that depends on $(t_1,t_2)$. In the above equation
we have put
\begin{equation}
\langle\Xi(s_1)\Xi(s_2)\rangle=
\int_0^\infty \! dt_1 \int_0^\infty \! dt_2
\langle\xi(t_1)\xi(t_2)\rangle e^{-s_1 t_1 + s_2 t_2}.
\label{eq:intehabo}
\end{equation}

In this work we consider an exponential correlator in Eqs.~\eqref{eq:corr3} and~\eqref{eq:corr4bis},
namely
\begin{equation}
	f(t) = \frac{1}{2\tau} e^{-|t|/\tau},\label{eq:expdec}
\end{equation}
which allows to solve the problem analytically.
In the following we call $\tau$ the memory time,
since it sets the time scale over which time correlations
of the noise decay.
Note that $\lim_{\tau\rightarrow 0}f(t)=\delta(t)$ in the distributional sense, therefore it is possible to interpolate between
the local and the nonlocal kernel by changing the value of $\tau$.
The Laplace transform of the dissipation kernel is
\begin{equation}
	\Gamma(s)=\frac{\gamma}{(\tau s + 1)}.\label{eq:24}
\end{equation} 
Moreover, a quick calculation shows that 
\begin{equation}
	\langle\Xi(s_1)\Xi(s_2)\rangle=	
	 \frac{ \mathcal{D} (2+\tau s_1 + \tau s_2)}
	 {(s_1 + s_2)(1+\tau s_1)(1+\tau s_2)}.
	\label{eq:intehabo2}
\end{equation}

\subsection{Thermalization}
Firstly, we examine the thermalization of the heavy particle, which consists in the loss of information about the
initial condition and in the equilibration of its
kinetic energy with the bath, $\langle E\rangle=T/2$
as required by the Equipartition Theorem. 
From Eqs.~\eqref{eq:17},~\eqref{eq:solution1aam} 
and~\eqref{eq:24}, considering that the zeroes of  $s+\Gamma(s)$ are the solutions of the equation
$s + \tau s^2 + \gamma =0$, we get by a straightforward application of the residues theorem
\begin{equation}
	\langle p(t) \rangle = p_0 \frac{e^{-\frac{t(1+A)}{2\tau}}(-1+A)}{2A} + 
	p_0 \frac{e^{-\frac{t(1-A)}{2\tau}}(1+A)}{2A},\label{eq:exact1}
\end{equation}
with $A\equiv \sqrt{1-4\gamma\tau}$. 

Before moving to
the thermalization time we pause on the
early, pre-thermalization 
behavior of  $\langle p(t) \rangle$,
in order to emphasize the differences between the motion 
with and without memory.
In order to simplify the discussion we assume that
$\tau\ll1/\gamma$ where $1/\gamma$ represents the
thermalization time for processes without memory.
For $t\ll\tau$ we get
\begin{equation}
	\langle p(t) \rangle  = p_0\left(1- \frac{\gamma  t^2}{2\tau}\right),~~~t\ll\tau_\mathrm{mem},\label{eq:app1}
\end{equation}
while for $\tau\ll t\ll 1/\gamma$ we get
\begin{equation}
	\langle p(t)  \rangle  =   p_0\left(1- \gamma t\right),\label{eq:app1nom}
\end{equation}
The memory changes the pre-thermalization 
evolution of $\langle p(t)  \rangle$ 
from linear to quadratic; 
in particular, this implies that the thermalization of the particle in a bath with memory
is slower than the one in a bath without memory 
with the same drag coefficient, $\gamma$. 
This result is confirmed by the calculation of the
thermalization time that makes use of the full result
\eqref{eq:exact1}.

\begin{figure}[t!]
	\begin{center}
		\includegraphics[scale=0.3]{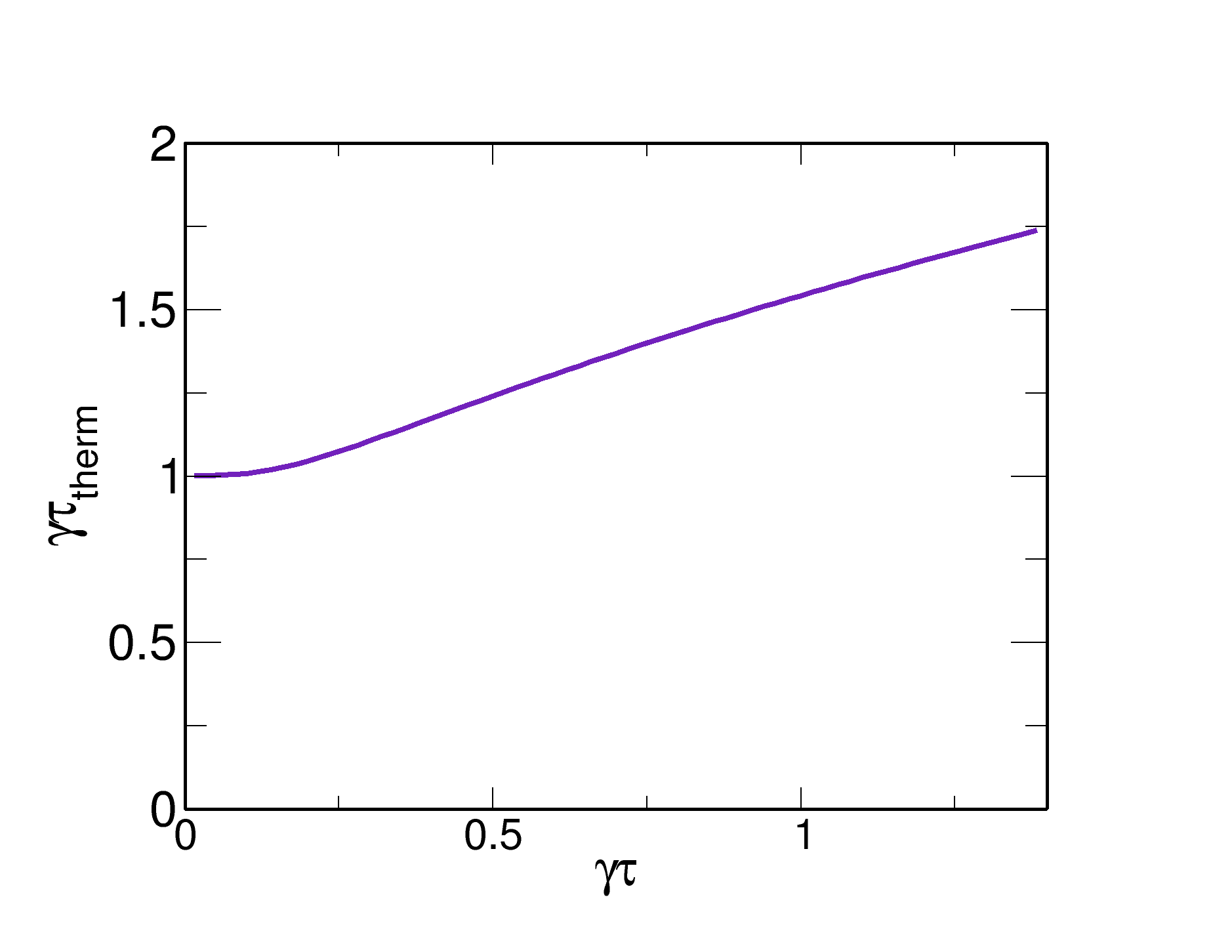}
	\end{center}
	\caption{\label{Fig:therm}Thermalization time versus memory time, both measured in units of $1/\gamma$.}
\end{figure}

We define the thermalization time, $\tau_\mathrm{therm}$, such that 
$\langle p(\tau_\mathrm{therm}) \rangle=p_0/e$
with $\langle p(t) \rangle$ given by Eq.~\eqref{eq:exact1}.
The results of this calculation are shown in 
Fig.~\ref{Fig:therm} where we plot
$\tau_\mathrm{therm}$ 
as a function of the memory time, both measured in units of $1/\gamma$.
Thermalization time increases with $\tau$ in agreement with the discussion above;
the quantitative effect of memory is negligible for $\gamma\tau\ll1$, but becomes substantial $\approx20\%$
already for $\gamma\tau\approx 0.5$.

The qualitative behavior of the thermalization time 
can be understood in simple terms. In fact, thermalization
implies the loss of the correlation with the initial
condition: the heavy particle equilibrates with the
bath regardless of its initial momentum distribution.
If the thermal noise of the bath has memory, the
momentum evolution from time $t$ to $t+\Delta t$
does not delete totally the information of $p(t)$
because of the correlations of the noise. Therefore,
it is natural to expect that the loss of the information about the
initalization requires more time.

\subsection{Momentum broadening}

Next we turn to the momentum broadening, $\sigma_p$ in Eq.~\eqref{eq:18}. The starting point is the computation 
of $\mathcal{J}$
in Eq.~\eqref{eq:solution1anna} with correlator given by
Eq.~\eqref{eq:intehabo2}.
We get
\begin{eqnarray}
	\sigma_p &=& \frac{{\cal D}}{2\gamma(4\gamma\tau-1)}e^{-\frac{t(1+A)}{\tau}}
	\nonumber\\
	&&\times
	\left[1-2\gamma\tau - 4\gamma\tau e^{\frac{t A }{\tau}} - A  -2A^2 e^{\frac{t(1+A)}{\tau}}\right.\nonumber\\
	&&\left.
	+(1-2\gamma\tau + A)e^{\frac{2t A }{\tau}}\right],\label{eq:marta2}
\end{eqnarray}
where $A$ has been defined right after Eq.~\eqref{eq:exact1}. 
Although Eq.~\eqref{eq:marta2} is exact, it is quite 
cumbersome, therefore 
it is convenient to analyze a few limit situations in which
the result  \eqref{eq:marta2} is manageable; after that,
we will present the full  result \eqref{eq:marta2} 
computed numerically. As in the previous subsection,
for the sake of illustration we assume that $\tau\ll 1/\gamma$:
this assumption will be removed in the full numerical
calculation, see Fig.~\ref{Fig:dnr}. 
From the results presented in the previous
subsection in this regime we have $\tau_\mathrm{therm} \approx
1/\gamma$, see Fig.~\ref{Fig:therm}.

The dissipative force becomes relevant on the time scale
$t\approx  \tau_\mathrm{therm}\approx 1/\gamma$, thus for
$t\gamma\ll1$ we can ignore it and
put $\gamma=0$ in  Eq.~\eqref{eq:marta2}. We get
	\begin{equation}
		\sigma_p = \mathcal{D}
		(2t-2\tau+2\tau e^{-t/\tau}),~~~t\ll 1/\gamma.
		\label{eq:nodragLIMIT654}
	\end{equation}
In the early stage $t\ll\tau$ the above result gives
	\begin{eqnarray}
		\sigma_p &\approx& \frac{\mathcal{D}t^2}{\tau},~~~
		t\ll\tau
		\ll 1/\gamma,
		\label{eq:nodragLIMIT654_bis}
\\
\sigma_p&\approx&2\mathcal{D}t,~~~
\tau \ll	t\ll
1/\gamma.\label{eq:nodragLIMIT654_bisA}
\end{eqnarray}
In particular, the result \eqref{eq:nodragLIMIT654_bisA}
agrees with that we would find for the Brownian motion
without memory. We notice that the effect of a finite
memory time is to slow down the momentum broadening
of the heavy particle, changing the early time evolution
from linear to quadratic.

For late times $t\gg 1/\gamma$, $t\gg\tau$, the memory
with the exponential kernel 
has no effect on the equilibration value of momentum broadening:
in fact, 
in the asymptotic limit $\gamma t\gg 1$ we get 
from Eq.~\eqref{eq:marta2}
\begin{eqnarray}
	\sigma_p &\asymp&   \frac{{\cal D}}{\gamma},
	\label{eq:asymp_expo_uuu_888}
\end{eqnarray}
in agreement with the standard result for the Brownian motion.

\section{Langevin equation with memory: numerical implementation
\label{sec:laimple}}
In this work, we solve the Langevin equation
for the heavy quarks in a bath
with a colored noise; the latter has correlations at different
times. In order to generate this noise we introduce an
ancillary stocastic process, $h(t)$, that evolves
simultaneously to (and independently of) the heavy quarks,
built up in such a way its correlator at different times
does not vanish. 
In this section we firstly define the ancillary process and specify
its correlations at different times; then, we formulate the
Langevin equation where the heavy quark is coupled to $h(t)$
and discuss the numerical discretization scheme
adopted in the calculations.

\subsection{Ancillary stochastic process}
Let us consider the stochastic process $a$ that satisfies
the Langevin equation
\begin{equation}
\frac{da}{dt} = -\alpha a + \alpha\xi,
\label{eq:xiacn1}
\end{equation}
where  
$\xi$ is a gaussian white noise,
\begin{eqnarray}
&&\langle \xi\rangle = 0,\\
&&\langle \xi(t)\xi(t^\prime)\rangle = \frac{1}{\alpha}
\delta(t-t^\prime); \label{eq:xiacn2}
\end{eqnarray}
$a$ is assumed to be dimensionless, so
the $1/\alpha$ in  \eqref{eq:xiacn2} is put to balance
the time dimension carried by the $\delta-$function because
$\xi$ is dimensionless too. 
The formal solution of Eq.~\eqref{eq:xiacn1} is given by
\begin{equation}
a(t) = a_0e^{-\alpha t}+
e^{-\alpha t} \int_0^t dt_1~\alpha\xi(t_1)e^{\alpha t_1},
\label{eq:xiacn3}
\end{equation}
where $a_0=a(t=0)$. Clearly we have
\begin{equation}
\langle a(t) \rangle = a_0e^{-\alpha t}.\label{eq:xiacn7}
\end{equation} 

We define the fluctuating field
\begin{equation}
h(t) \equiv a(t) - \langle a(t) \rangle.
\label{eq:deffluct77}
\end{equation} 
This satisfies Eq.~\eqref{eq:xiacn1} with $h(t=0)=0$, that we
rewrite for the sake of future reference:
\begin{equation}
	\frac{dh}{dt} = -\alpha h + \alpha\xi.
	\label{eq:xiacn1b}
\end{equation}
We baptize $h$ as the ancillary process, because we use
it as an additional stocastic process to generate the
colored noise for the Langevin equation of the
heavy quarks, see below.

From the very definition of $h$ it is easy to see that
$\langle h(t) \rangle =0$. 
Instead the correlator of $h$ at different times is 
not a $\delta-$function: it can be obtained easily
from Eq.~\eqref{eq:xiacn3}, namely
\begin{equation}
\langle h(t) h(t^\prime)\rangle
=
\frac{e^{-\alpha|t-t^\prime| }- e^{-\alpha(t+t^\prime)}}{2}.
\label{eq:stomem3}
\end{equation}
Therefore, $h$ is a process with memory and it 
can be used in any Langevin equation.
From 
Eq.~\eqref{eq:stomem3} it is obvious that asymptotically 
\begin{equation}
\langle h(t) h(t^\prime)\rangle \approx \frac{e^{-\alpha|t-t^\prime|}}{2},
\label{eq:stomem6}
\end{equation}
namely correlations are washed out on the time scale
$1/\alpha \equiv \tau$.
Note that for $\alpha\rightarrow+\infty$ we get,
from Eq.~\eqref{eq:stomem3},
\begin{equation}
\alpha\langle h(t) h(t^\prime)\rangle
	\approx
	 \delta(t-t^\prime),
	\label{eq:stomem3b}
\end{equation}
that is the process $h$ becomes a standard white noise in the
limit $\tau\rightarrow 0$, as expected.

Before going ahead, it is useful to comment on
Eq.~\eqref{eq:stomem3}: 
we note that the correlator is not a function of $t-t^\prime$
but of $t$ and $t^\prime$ separately. 
It is convenient
to fix $t^\prime$ and analyze the correlator for 
$t\geq t^\prime$. The addendum $A\equiv 
e^{-\alpha(t+t^\prime)}$
lowers the value of the correlator; 
on the other hand, $A$ is suppressed
when $t^\prime=O(1/\alpha)$. That is, correlations of $h$
 develop substantially on a time scale 
 $t\approx\tau=1/\alpha$
necessary to suppress $A$.
After this transient regime, time correlations are approximately
given by Eq.~\eqref{eq:stomem6} which have also the property
to be invariant under time translations. 
Therefore, the process \eqref{eq:xiacn1b} describes a noise
that needs a time $\approx\tau$ to develop
memory: after the system enters in this regime, the 
correlations of $h$ at different times are approximately
invariant under time translation, and decay on a time scale
$\approx\tau$.

Numerically Eq.~\eqref{eq:xiacn1b} can be discretized in the
usual manner by the replacements 
\begin{eqnarray}
&&\delta(t-t^\prime) \rightarrow \frac{\delta_{t,t^\prime}}
{\Delta t},
\label{eq:pop1}\\
&&\xi(t) =\zeta(t)\sqrt{\frac{1}{\alpha\Delta t}},
\label{eq:pop2}
\end{eqnarray}
where $\Delta t$ corresponds to the discrete time step
implemented in the numerical calculation. With these
we have 
\begin{equation}
\Delta h = -\alpha h \Delta t +
\sqrt{\alpha}\zeta(t) \sqrt{ \Delta t}.
\label{eq:ppp7}
\end{equation}
$\zeta(t)$ will be implemented as a white noise with variance
equal to one.

\subsection{Application to the Langevin equation}
Next we turn to the solution of the 
Langevin equation,
\eqref{eq:L1}, for heavy quarks in the relativistic
limit. 
We assume the FDT in the
relativistic limit, namely 
\begin{equation}
		\gamma(t,s) = \frac{1}{E T}\langle
		\eta(t) \eta(s)
		\rangle,
		\label{eq:FDT_1r}
\end{equation}
with $E=\sqrt{\bm p^2 + M^2}$. 
Moreover,
the process $\eta(t)$ in \eqref{eq:L1} is assumed to 
satisfy $\langle\eta\rangle=0$ and
\begin{equation}
	\langle\eta(t_1)\eta(t_2)\rangle =  \frac{2{\cal D}}
	{2\tau} g(t_1-t_2),
	\label{eq:sbd}
\end{equation}
where $\tau$ 
is the memory time and $g$ is a dimensionless function that 
defines the correlation of the noise; 
for simplicity we assume it satisfies $g(0)=1$.
In the case of a Markov process
$\tau\rightarrow 0$ and
$g$ has to satisfy the condition
\begin{equation}
	\frac{1}{2\tau}g(t_1-t_2) \rightarrow\delta(t_1-t_2)
	.\label{eq:attana}
\end{equation}

In this work we generate the noise $\eta$ by means of the
ancillary process $h$ introduced in the previous subsection
with $\alpha=1/\tau$. More precisely
we assume that 
\begin{eqnarray}
\frac{1}{2\tau}g(t-t^\prime)
&=&\frac{1}{\tau}\langle
h(t) h(t^\prime)\rangle,
\label{eq:daref789}
\end{eqnarray}
see Eq.~\eqref{eq:stomem3}. 
Therefore, in Eq.~\eqref{eq:L1} we put
\begin{equation}
\eta(t) = \sqrt{\frac{2\mathcal{D}}{\tau }}h(t),
\label{eq:rescaling}
\end{equation}
see also Eq.~\eqref{eq:sbd}.
 Using Eqs.~\eqref{eq:stomem3b}
and~\eqref{eq:sbd}  we note that 
in the limit 
$\tau\rightarrow 0$ we get 
\begin{equation}
\langle\eta(t)\eta(t^\prime)\rangle =
 2\mathcal{D}\delta(t-t^\prime),
 \label{eq:uncorrelatednoise555}
\end{equation}
namely we recover the time correlation of the
standard Brownian motion.

We have noted
that the ancillary process \eqref{eq:xiacn1b} develops
substantial correlations after an initial
transient stage that lasts for
$t^\prime\approx \tau $; 
after this transient, the correlator is approximately 
invariant under time translation and decays exponentially on
the time scale $\tau $. 
We call this stage as the exponential decay regime.
In the numerical calculations
we start the ancillary process and let it run 
up to some time $t_0$, leaving heavy quarks frozen in momentum
and coordinate space; then, when the noise is in the 
exponential decay regime, we 
start the evolution of the heavy quarks including
their interaction with the noise itself.
In this regime the correlator takes the form
	\begin{equation}
		\langle\eta(t)\eta(t^\prime)\rangle =
		2\mathcal{D}
		\frac{e^{-|t-t^\prime|/\tau  }}
		{2\tau }.
		\label{eq:correlatedMODELsimple}
	\end{equation}

With the rescaling \eqref{eq:rescaling} 
the Langevin equation~\eqref{eq:L1} reads
\begin{equation}
	\frac{dp}{dt} = -\int_{t_0}^t \gamma(t,s)p(s) ds + \sqrt{\frac{2\mathcal{D}}{\tau }}h(t),
	\label{eq:L1b}
\end{equation}
that we solve for $t>t_0$.
In \eqref{eq:L1b} we put
\begin{equation}
	\gamma(t,s) = \frac{	2\mathcal{D}}{ET}
	\frac{e^{-|t-s|/\tau  }}
	{2\tau },\label{eq:FDtheor_implem}
\end{equation}
in agreement with the relativistic form of the FDT.
The time-discretized version of Eq.~\eqref{eq:L1b}
is given by
\begin{equation}
\Delta p = -\Delta t\int_{t_0}^t \gamma(t,s)p(s) ds +
\sqrt{\frac{2\mathcal{D}}{\tau }}h(t)\Delta t.
\label{eq:L1bis}
\end{equation}
Note that differently from the Markov process,
the stochastic term in \eqref{eq:L1bis} does not come
with the $\sqrt{\Delta t}$: this is so because now the
noise is not a Wiener process due to the memory.
We implement a simple iterative scheme to solve
\eqref{eq:L1bis}, namely
\begin{eqnarray}
	p(t) &=& p(t-\Delta t) 
	-\Delta t\sum_{n=0}^{N} \gamma(t,s_n)p(s_n)   \Delta t
	\nonumber\\
	&&
	+
	\sqrt{\frac{2\mathcal{D}}{\tau }}h(t)\Delta t,
	\label{eq:L1bis3aaa}
\end{eqnarray}
with $s_0=t_0$, $s_{N-1}=t-\Delta t$ 
and
$s_N = t$.
This scheme leads to the self-consistent solution
\begin{eqnarray}
	p(t)
	\left[ 1+(\Delta t)^2\gamma(t,t)\right]
	 &=& p(t-\Delta t) \nonumber\\
	 &&
	-(\Delta t)^2\sum_{n=0}^{N-1} \gamma(t,s_n)p(s_n)    
	\nonumber\\
	&&
	+
	\sqrt{\frac{2\mathcal{D}}{\tau }}h(t)\Delta t.
	\label{eq:L1bis3}
\end{eqnarray}

Equations  \eqref{eq:ppp7} and \eqref{eq:L1bis} allow us to
implement the momentum evolution 
of the heavy quark in a bath with a colored noise.
$h$ in Eq.~\eqref{eq:L1bis} is given
by the solution of the ancillary Langevin equation 
\eqref{eq:ppp7}. This means that at each time step
one has to solve \eqref{eq:ppp7} 
[with the initial condition
$h(t=0)=0$] and \eqref{eq:L1bis} 
simultaneously.  This procedure is different from the one
adopted in the literature in the case of a Markov process,
in which the white noise would be extracted randomly from a
Gaussian distribution at each time step.

For the numerical implementation of the 3-dimensional Langevin equation, we are solving equations  \eqref{eq:ppp7} and \eqref{eq:L1bis} simultaneously for the 3-components 
($h_x$,$h_y$, $h_z$, $p_x$, $p_y$, $p_z$) to study HQ momentum 
evolution coupled with the coordinate evolution
\begin{equation}
 dr_i=\frac{p_i}{E×}dt
\end{equation}
where $dr_i$ is the shift of the coordinate  in each time step $dt$.
$E$ and $p$ are the energy and momentum of the heavy quark respectively.

The HQ transport coefficients are computed as 
follows.
At high $T$ we model the thermalized bath by
a QGP made of massless quarks and gluons, and use the pQCD 
kinetic coefficients for the processes  
$c \ell \rightarrow c \ell$,
where $\ell$ denotes either 
a massless gluon or a quark in the bath. 
The diffusion coefficients in this case
are well known and can be found in the literature, see
for example \cite{comb, Svetitsky:1987gq}. 
The squared invariant scattering amplitudes are the Combridge ones, 
that includes $s$, $u$ and $t$ channel and their interferences terms. The 
infrared 
divergence associated with the $t$-channel diagrams
is 
screened by the Debye mass, 
$m_D = g(T)T$. For more details  
see earlier works \cite{Das:2015ana,Scardina:2017ipo}. 
On the other hand,
at low $T$ we model the bath with a gas of quarks and gluons quasiparticles, using
the so-called 
quasi-particle models (QPMs);
in these models the quasiparticle masses 
are tuned in order to reproduce Lattice QCD thermodynamics~\cite{Plumari:2011mk,Berrehrah:2013mua}.The QPMs  account effectively for
the non-perturbative 
effects for $T$
close to the 
quark-hadron transition 
temperature, $T_c$.  
The main feature of the QPM  is that the
effective coupling is significantly stronger than the one of pQCD near $T_c$, which enhances the HQ-bulk inteactions.
We have evaluated the diffusion coefficient within the QPM  starting from the effective coupling with massive quarks and gluons. 
For details we refer to earlier works~\cite{Das:2015ana, Scardina:2017ipo}.

\section{Results}

\subsection{Non-relativistic check}

\begin{figure}[t!]
	\begin{center}
		\includegraphics[scale=0.3]{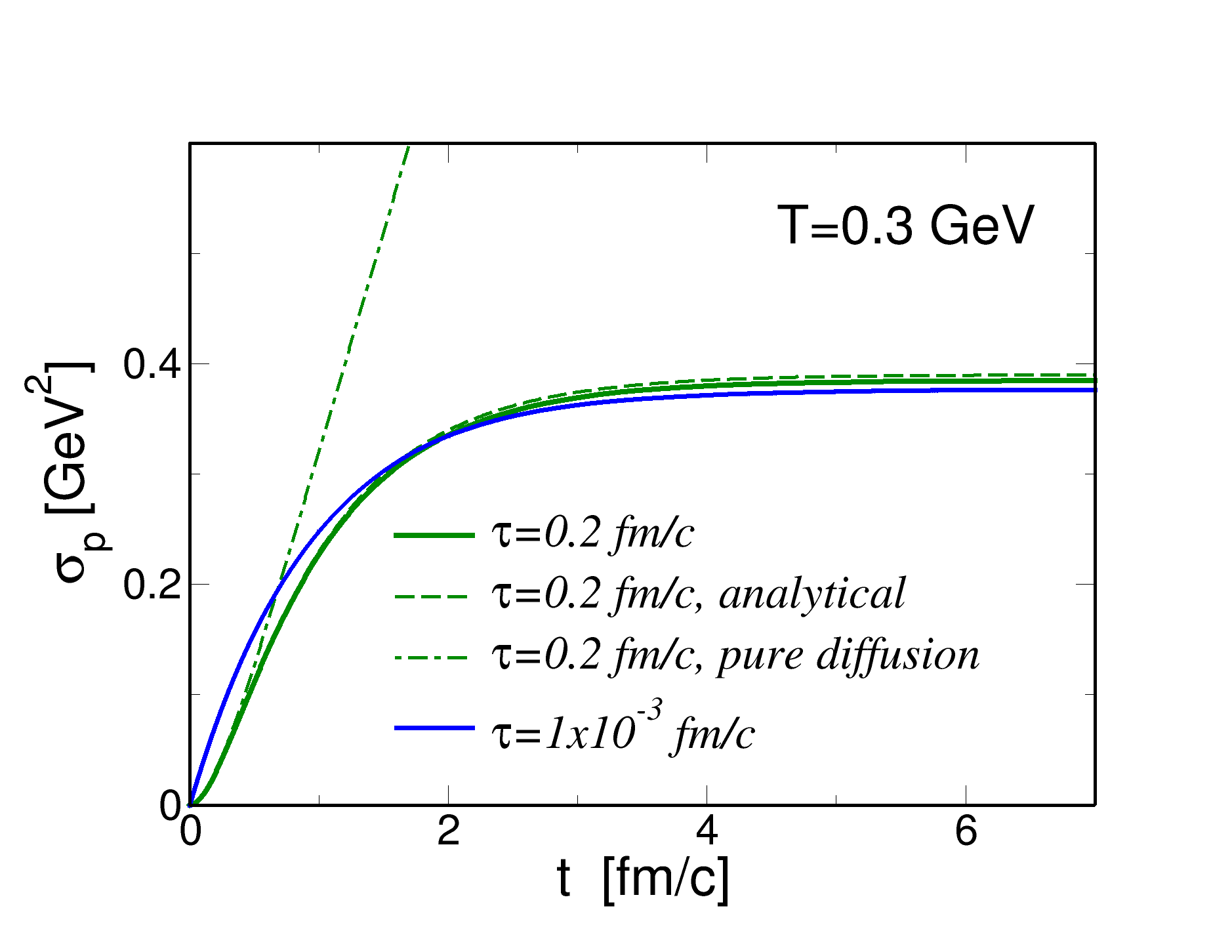}
	\end{center}
	\caption{\label{Fig:dnr}
		$\sigma_p$ versus time for a one-dimensional motion
		of charm quarks in the nonrelativistic limit,
		see Eq.~\eqref{eq:varrhDEDO}. 
	$\sigma_p$ is	defined in Eq.~\eqref{eq:18}.
	We have considered  
		 two values of the memory time, $\tau$.
	 Analytical results correspond to
 Eq.~\eqref{eq:marta2}.}
\end{figure} 

In order to check the discretization scheme of the Langevin
equation with correlated noise \eqref{eq:L1bis},
we plot $\sigma_p$ versus time
in Fig.~\ref{Fig:dnr}; the definition of  $\sigma_p$
is given in Eq.~\eqref{eq:18},
and the Langevin equation  has been solved for a one-dimensional
motion of charm quarks.
We have solved the equation in the nonrelativistic limit
as explained in section \ref{sec:iianalyt}, that corresponds
to replace the kinetic energy with the mass of the quark
in the Fluctuation-Dissipation theorem, 
see Eq.~\eqref{eq:varrhDEDO};
moreover, we have used a constant diffusion coefficient
$\mathcal{D}=0.2$ GeV$^2$/fm
for illustrative purposes only. 
Here $T$ referes to the bath temeprature.
The green and orange solid lines correspond to
$\tau=0.2$ and $\tau=0.01$ fm/c.

For comparison, in the same figure we show by dashed lines
the analytical result \eqref{eq:marta2},
that is valid in the nonrelativistic limit: the agreement 
between the two results is excellent, showing that our
numerical scheme works properly.
We notice that the memory slows down the evolution
of $\sigma_p$ as anticipated in section \ref{sec:iianalyt}.
Moreover, for $\tau=0.2$ fm/c
we notice that the diffusive motion is characterized by
the initial nonlinear increase of $\sigma_p$
that turns  into  a linear regime before the drag force becomes 
relevant and leads to the thermalization of the heavy quark.
In the case of the smaller $\tau$ the charm quark 
enters the linear regime immediately, then equilibrates 
with the medium.

\subsection{Transverse momentum broadening}
In this subsection we analyze momentum broadening of heavy quarks
in a hot medium; analogously to the previous
section we define 
\begin{equation}
\sigma_p = \langle 
(p_T - \langle p_T \rangle)^2
\rangle,
\label{eq:sigmap_pT}
\end{equation} 
where $p_T=\sqrt{p_x^2 + p_y^2}$ is the transverse momentum.
 
\begin{figure}[t!]
	\begin{center}
		\includegraphics[scale=0.3]{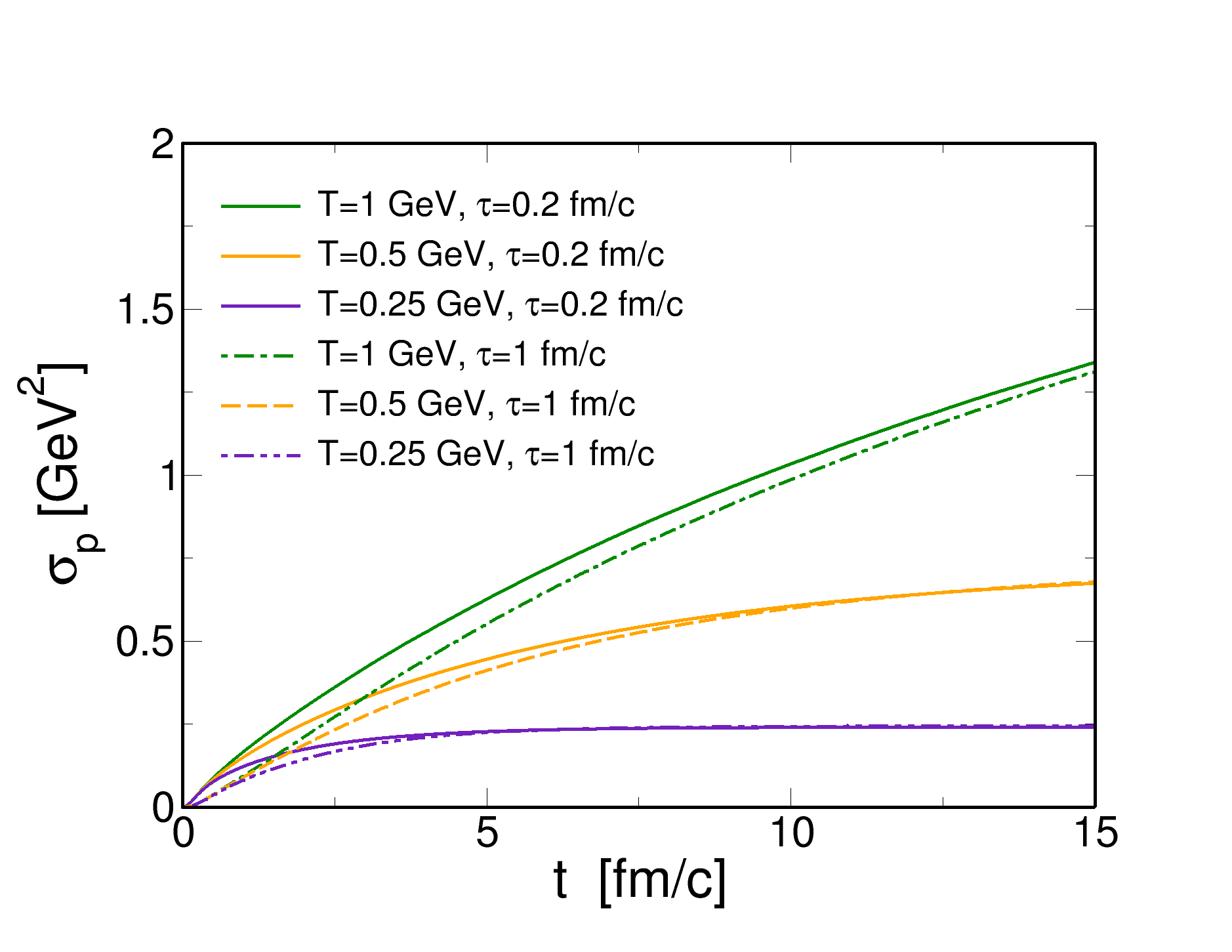}
	\end{center}
	\caption{\label{Fig:relattau}
	$\sigma_p$ for charm quarks, 
	defined in Eq.~\eqref{eq:sigmap_pT},
	versus time, for three values of the
temperature. The solid lines correspond to $\tau=0.2$ fm/c
while the dashed, dot-dashed and dot-dot-dashed stand for
$\tau=1$ fm/c.}
\end{figure} 

In Fig.~\ref{Fig:relattau} we plot $\sigma_p$ versus
time for three temperatures and two values of $\tau$,
namely $\tau=0.2$ fm/c (solid lines) and $\tau=1$ fm/c
(broken lines).
Calculations correspond to the cham quark. 
For the three cases considered it is clear that increasing
the memory time results in the slowing down of
momentum broadening; the effect is more visible
at large temperature, where the drag force is less effective. 
At small temperature it is still possible to measure some
difference between the results with the two memory times
in the early evolution, then for time $t\geq 5$ fm/c the results
with and without memory coincide.

\begin{figure}[t!]
	\begin{center}
		\includegraphics[scale=0.3]{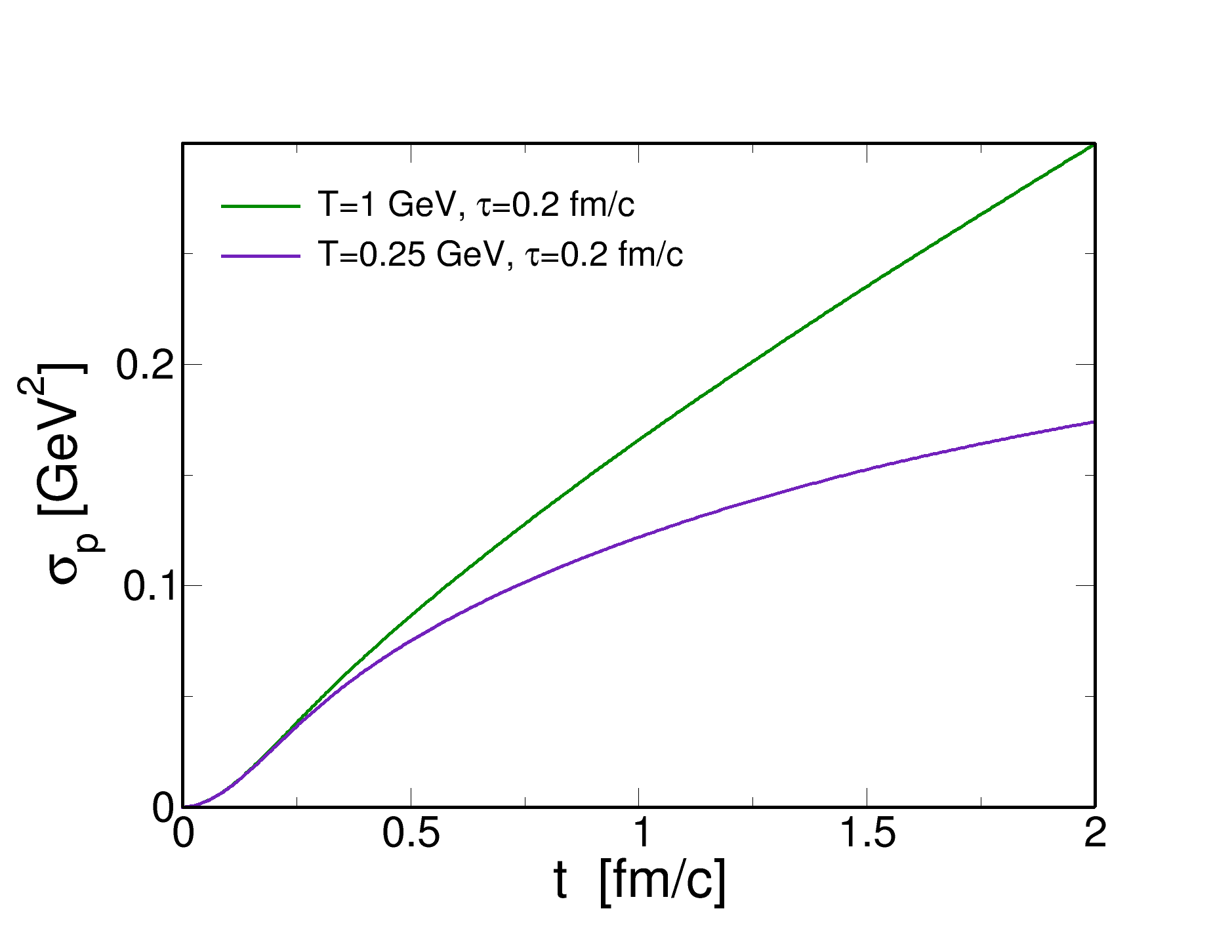}
	\end{center}
	\caption{\label{Fig:relattauea}
		$\sigma_p$ for charm quarks, 
		defined in Eq.~\eqref{eq:sigmap_pT},
		versus time and for $\tau=0.2$ fm/c. 
		}
\end{figure} 

In Fig.~\ref{Fig:relattauea} we plot a selection of the
results shown in Fig.~\ref{Fig:relattau} for $\tau=0.2$ fm/c,
zooming on  the early time evolution of $\sigma_p$.
We notice the nonlinear increase of $\sigma_p$,
induced by memory
in agreement with the discussion of section \ref{sec:iianalyt},
followed by a linear enhancement.
In this regime the charm quarks experience an almost diffusive
motion, in the sense that the energy loss due to the drag force
is still negligible.
Qualitatively the
different regimes appear also
for the small temperature case in the
figure: however, in this case the drag force is stronger so the 
linear regime lasts for a shorter fraction of the evolution,
then $\sigma_p$ bends and eventually saturates,
signaling the equilibration of the charm with the medium.

\subsection{Nuclear modification factor}
In this subection we analyze the modification factor $R_\mathrm{AA}$, defined as
\begin{equation}
R_\mathrm{AA}(p_T) = \frac{(dN/d^2p_T)_\mathrm{t}}
{(dN/d^2p_T)_\mathrm{FONLL}},
\label{eq:raadefc}
\end{equation}
where $(dN/d^2p_T)_\mathrm{t}$ denotes the spectrum
of charm quarks at time $t$ and 
$(dN/d^2p_T)_\mathrm{FONLL}$ denotes the spectrum
at the initialization time, 
To this end, at the formation time we assume the prompt spectrum obtained
within 
Fixed Order + Next-to-Leading Log (FONLL) QCD that 
reproduces the D-mesons spectra in $pp$ collisions after fragmentation~\cite{FONLL,Cacciari:2012ny}.  
\begin{equation}
	\left.\frac{dN}{d^2 p_T}\right|_\mathrm{prompt} = \frac{x_0}{(1 + x_{3}{p_{T}}^{x_1})^{x_2}};\label{eq:FFNLO}
\end{equation}
the parameters that we use in the calculations are $x_0=20.2837$, $x_1=1.95061$, $x_2=3.13695$ and $x_3=0.0751663$ for 
charm quarks;
the slope of the spectrum has been calibrated to a collision at $\sqrt{s}=5.02$ TeV.
$R_\mathrm{AA}(p_T)\neq 1$ implies that charm quarks
experience interactions with the gluon medium, causing a
change in their spectrum. Our motivation is to highlights 
the impact of memory on $R_\mathrm{AA}(p_T)$.

\begin{figure}[t!]
	\begin{center}
		\includegraphics[scale=0.27]{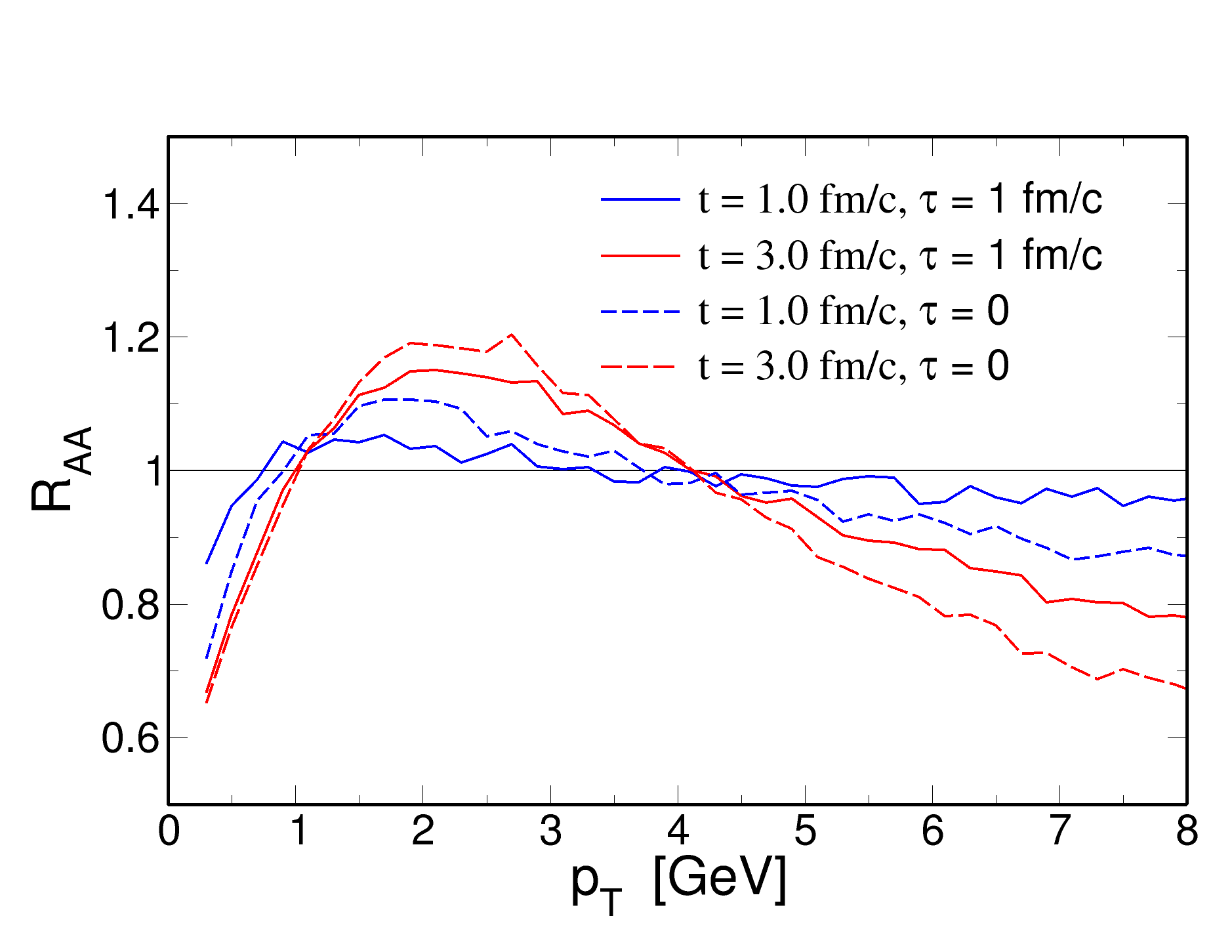}
	\end{center}
	\caption{\label{Fig:raavariet}
		$R_\mathrm{AA}$ at $t=1$ fm/c and $t=3$ fm/c,  for $\tau=0$ and $\tau=1$ fm/c , at $T=1$ GeV obtained within pQCD.
		}
\end{figure} 

In Fig.~\ref{Fig:raavariet} we plot $R_\mathrm{AA}$ 
versus $p_T$ , for two values
of the memory time $\tau$. 
The shape of $R_\mathrm{AA}$ for $T=1$ GeV and up to
$p_T\approx 5$ GeV is in agreement
with the one advertised in 
\cite{Ruggieri:2018rzi,Liu:2019lac,Liu:2020cpj}:
this is the result of the diffusion-dominated propagation
of the heavy quarks in the hot medium at high temeprature,
that effectively diffuses low momentum charm quarks to higher momentum states. This explains why
$R_\mathrm{AA}$ remains smaller than one up to 
$p_T\approx 2$ GeV.
For larger $p_T$ the energy loss is important and
particles migrate to lower momentum states.
At low temperature the drag force is
dominant in the whole range of $p_T$,
therefore the general tendency is that particles move 
to lower $p_T$ states so $R_\mathrm{AA}$ stays greater than one
up to $p_T\approx 2$ GeV.

\begin{figure}[t!]
	\begin{center}
		\includegraphics[scale=0.27]{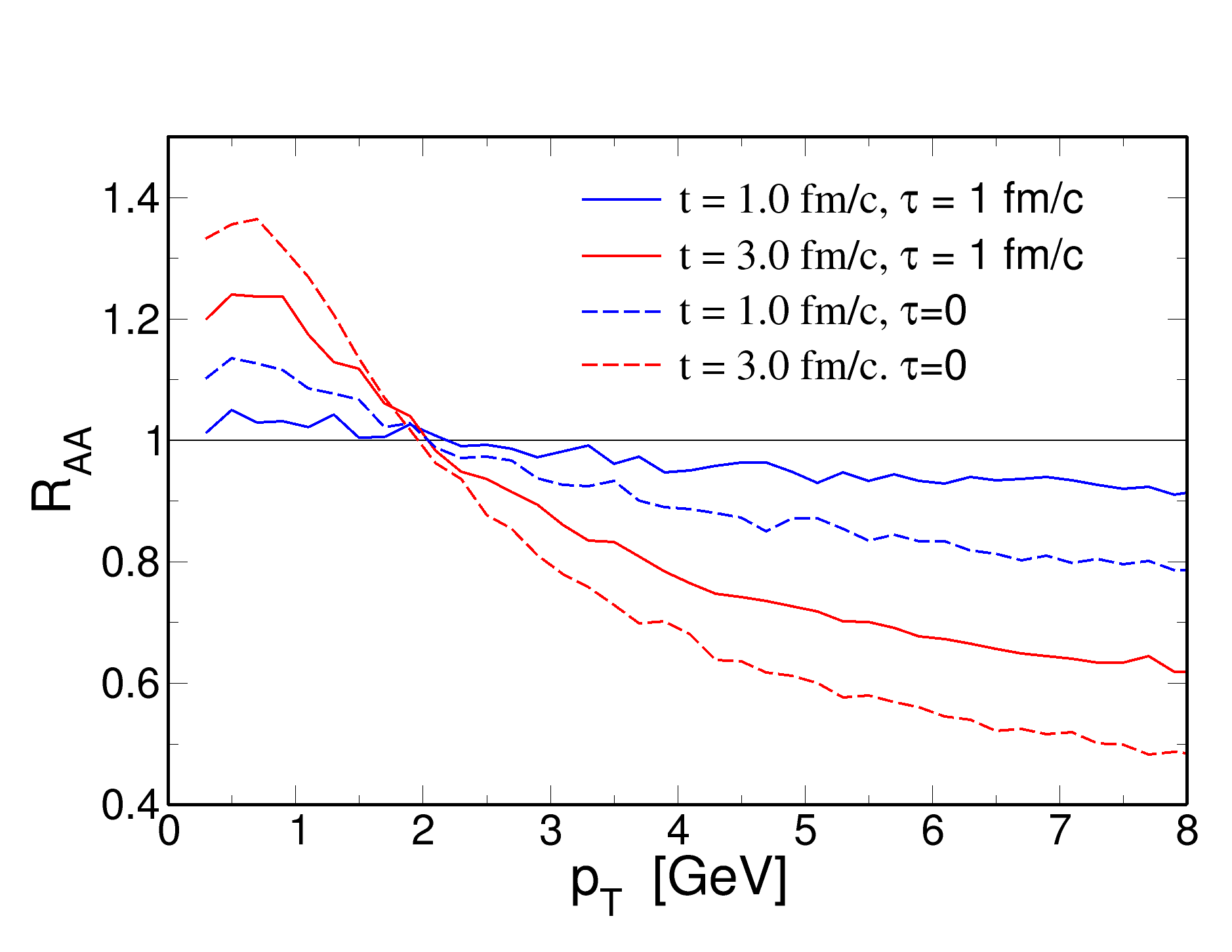}
	\end{center}
	\caption{\label{Fig:compa025}
		$R_\mathrm{AA}$ at $t=1$ and $t=3$ fm/c, 
		for $\tau=0$ and $\tau=1$ fm/c, at $T=0.25$ GeV
		obtained within QPM.
	}
\end{figure}

The effect of $\tau\neq 0$ is to slow down the formation of 
$R_\mathrm{AA}$. We notice a sizable impact of memory on 
the $R_\mathrm{AA}$. To make this poing clearer, in Fig.~\ref{Fig:compa025}
we show $R_\mathrm{AA}$ for $\tau=0$ and $\tau=1$ fm/c
at  $T=0.25$ GeV obtained within QPM. We notice the slower evolution of $R_\mathrm{AA}$ when $\tau\neq 0$.

\begin{figure}[t!]
	\begin{center}
		\includegraphics[scale=0.27]{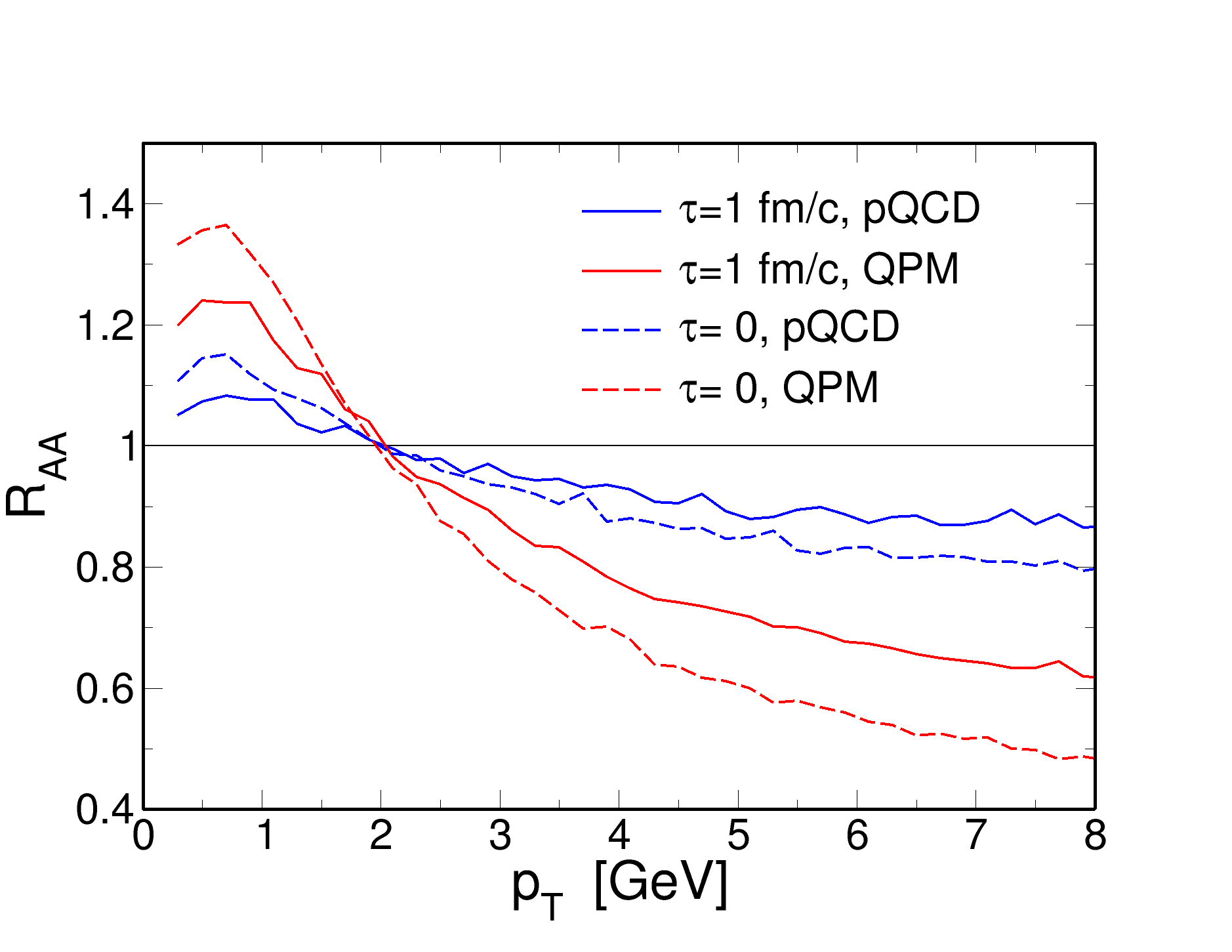}
	\end{center}
	\caption{\label{Fig:raaqpm}
		$R_\mathrm{AA}$ at  $t=3$ fm/c, 
		for $\tau=0$ and $\tau=1$ fm/c, at $T=0.25$ GeV.
		Red lines correspond to the results of the QPM 
		model, while blue lines stand for the pQCD calculations.
	}
\end{figure}

In Fig.~\ref{Fig:raaqpm} we plot $R_\mathrm{AA}$ 
at  $t=3$ fm/c, 
for $\tau=0$ and $\tau=1$ fm/c, at $T=0.25$ GeV.
Solid lines correspond to the results of the QPM 
model, while dashed lines stand for the pQCD calculations.
We notice that
$R_\mathrm{AA}$ obtained within the QPM model differs
considerably from the results of pQCD, which is in agreement
with expectations because the cross sections within the former
are enhanced with respect to those of the latter,
implying that diffusion and drag coefficients are larger.

The results in Fig.~\ref{Fig:raaqpm}
show that memory slows down the dynamics 
of the heavy quarks in the whole
$p_T$ range. In fact, $R_{AA}$ at large $p_T$ in the
process with memory stays above that without memory,
meaning that the large $p_T$ particles have lost less energy
during the evolution. 
In other words, memory slows down the energy loss.
Ergo, also thermalization in presence of
a time-correlated noise is retarded.
Similarly, the diffusion to low $p_T$
is slower when memory is present, since $R_{AA}$ in this case
remains lower than the one without memory.
These effects are more evident in the case of the QPM
while they are smaller (but still present) in the pQCD case.

The results discussed in this section 
allow us to discuss 
a potential impact that
memory might have on more sophisticated
phenomenological calculations.
As a matter of fact,  our results can be summarized by stating
that memory slows down the heavy quark dynamics. 
For example, the formation 
of $R_{AA}$ of heavy quarks 
is slower if the thermal fluctuations of the
bath are time correlated. Therefore, if the 
diffusion coefficient is tuned in order to reproduce the
experimental $R_{AA}$, then a larger coefficient is needed
when the bath has memory. 
The larger diffusion coefficient will then enhance the 
elliptic flow, $v_2$. The memory effect has the potential 
to alter the heavy quark $R_{AA}$-$v_2$ dynamics and can 
improve the simulations description of heavy quark $R_{AA}$-$v_2$ ~\cite{Das:2015ana}. 
Memory effect may affect other
observables like the heavy quark directed flow, particle correlations
and so on.
Within our present study it is
impossible to predict quantitatively this change due to our
simplifying assumptions; 
these simplifications are partly justified 
by the fact that it is the first time, to our knowledge,
that the effects of time correlated fluctuations on
heavy quark observables are computed.
A more detailed, quantitative study will be the 
subject of future investigations.

\section{Conclusions} 
We have studied the effects of a 
time-correlated thermal noise 
of a thermalized quark-gluon plasma 
on the energy loss and
the diffusion of heavy quarks. 
In this case the time correlation
of the thermal noise does not decay instantaneously,
as instead it is assumed  for the standard Brownian motion. 
We have considered a simple situation in which the time
correlations of the noise decay exponentially
on the time scale $\tau$, called the memory time, 
and treated $\tau$ as a free parameter.
We have considered 
an integro-differential Langevin equation for the heavy
quark momentum,
taking into account memory effects both in the
thermal noise and in the dissipative force.

We have computed several quantities that characterize
the dynamics of heavy quarks in the bath, namely the 
thermalization time and the transverse momentum broadening.
Then we turned to the nuclear modification factor, $R_{AA}$,
that we have computed using kinetic coefficients
from pQCD and quasiparticle models.
Our results suggest that
memory delays the dynamics of the heavy quarks in the
QGP: indeed, memory
slows down momentum broadening as well as
the formation of $R_{AA}$, 
retards the energy loss and thus 
increases the thermalization time.

Our work could be of interest 
for the phenomenology
of heavy quarks in the QGP: in fact,  
the slower evolution of  $R_{AA}$ would require the use
of larger diffusion coefficients in phenomenological
calculations, for
reproducing the experimental $R_{AA}$ and this in turn would
require stronger interactions of the heavy quarks with the
bulk, potentially leading to a larger $v_2$. 
The investigation of this point requires a refined study
with realistic initial conditions, including the 
initial geometry of the fireball, and will be matter
for future studies.
In addition to this problem, it is of a certain interest
to analyze processes with a long tail memory, in which
the time correlations of the noise do not decay
exponentially but as power laws. 
These processes can be
generated via Langevin equations with
fractional derivatives \cite{fc5,fc6};
this problem will also be the subject of future studies.

\begin{acknowledgements}
	M. R. acknowledges John Petrucci for inspiration
	and Marco Frasca for discussions on the results of
	section \ref{sec:iianalyt}.
	M. R. is supported by the National Science Foundation of China (Grants No.11805087 and No. 11875153). S.K.D. acknowledges 
	Amaresh Jaiswal for useful discussions. 
	S.K.D. acknowledges the support from DAE-BRNS, India, Project No. 57/14/02/2021-BRNS.
\end{acknowledgements}

\end{document}